\documentclass[twoside,fleqn]{article}
\usepackage{espcrc2}
\usepackage{graphics}
\usepackage{graphicx}

\newcommand{\AmS}{{\protect\the\textfont2
  A\kern-.1667em\lower.5ex\hbox{M}\kern-.125emS}}

\title{Experimental moments of the nucleon structure function $F_2$}

\author{M. Osipenko\address{Istituto Nazionale di Fisica Nucleare, 
        Sezione di Genova, Genoa, Italy 16146}, %
        W. Melnitchouk\address{Jefferson Lab, Newport News, Virginia 23606},
	S. Simula\address{Istituto Nazionale di Fisica Nucleare, Sezione Roma III,
	Roma, Italy 00146},
	S. Kulagin\address{Institute for Nuclear Research of Russian Academy of Science,
	Moscow, Russia 117312},
	G. Ricco\address{Universit\`a di Genova, Genoa, Italy 16146}
	and CLAS Collaboration
	}
       
\begin{document}

\begin{abstract}
Experimental data on the $F_2$ structure functions of the proton
and deuteron, including recent results from CLAS at Jefferson Lab,
have been used to construct their $n \leq 12$ moments.
A comprehensive analysis of the moments in terms of the operator
product expansion has been performed to separate the moments into
leading and higher twist contributions.
Particular attention was paid to the issue of nuclear corrections
in the deuteron, when extracting the neutron moments from data.
The difference between the proton and neutron moments was compared
directly with lattice QCD simulations.
Combining leading twist moments of the neutron and proton we
found the $d/u$ ratio at $x\to 1$ approaching 0,
although the precision of the data did not allow to
exclude the 1/5 value.
The higher twist components of the proton and neutron moments
suggest that multi-parton correlations are isospin independent.
\end{abstract}

\maketitle

The Operator Product Expansion (OPE) is a powerful tool in QCD 
which allows measurable moments of hadronic structure functions to
be related to series expansions of the moments in terms of twists.
The first term in the series, corresponding to Leading Twist (LT),
reflects the physics of asymptotic freedom, and is determined by
single-parton distributions in the hadron.
Subsequent terms in the series, or Higher Twists (HT), describe
interactions between partons, or multi-parton correlations.
The determination of the HTs is considerably more challenging,
both experimentally and theoretically.

We analyzed data on experimentally extracted moments of the
proton and deuteron structure functions $F_2$~\cite{osipenko_f2pd}
to separate LT and HT terms.
The $n$-th moment of the $F_2$ structure function, including
LT and HT contributions, can be written:
\begin{eqnarray}
&& M_n (Q^2) = LT_n(\alpha_S) \nonumber \\
&& + \sum_{\tau=4,6} a_n^\tau
\Biggl(\frac{\alpha_S(Q^2)}{\alpha_S(\mu^2)}\Biggr)^{\gamma_n^\tau}
\Biggl(\frac{\mu^2}{Q^2}\Biggr)^{\frac{\tau-2}{2}} ~,
\end{eqnarray}
\noindent where $LT_n$ is the leading, twist-2 moment, $\alpha_S$
is the running coupling constant,
$\mu^2$ is an arbitrary scale (taken to be 10 (GeV/c)$^2$),
$a_n^\tau$ is the matrix element of corresponding QCD operators,
$\gamma_n^\tau$ is the anomalous dimension
and $\tau$ is the order of the twist.

\begin{figure}[htb]
\vspace{9pt}
\includegraphics[bb=2cm 4cm 22cm 24cm, scale=0.4]{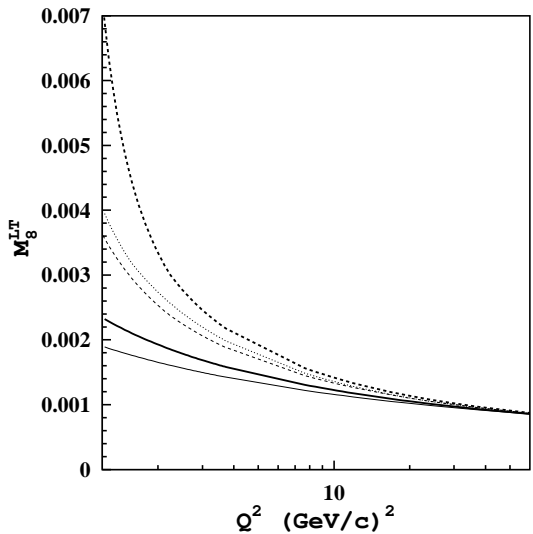}
\caption{$n=8$ LT moment calculated to different pQCD accuracy:
thin solid line - LO,
thin dashed line - NLO,
thin dotted line - NNLO,
thick solid line - LL
thick dashed line NLL.}
\label{fig:pqcd_orders}
\end{figure}

The separation of the LT from the complete series is to some extent 
dependent on the order to which one calculates the LT $Q^2$-evolution.
In Fig.~\ref{fig:pqcd_orders} we compare the $n=8$ moment of the
LT term calculated at fixed orders in pQCD: Leading Order (LO),
Next-to-Leading Order (NLO) and Next-to-Next-to-Leading Order (NNLO);
and by using resummation of soft gluon emission (SGR)~\cite{SGR}:
Leading Log (LL) and Next-to-Leading Log (NLL).
The fixed order calculations appear to converge at NLO\footnote{
	The difference between NLO and NNLO is small with respect
	to uncertainties of the data}.
However, at fixed pQCD order the logarithmic precision of the LT
term deteriorates the closer one gets to $x = 1$.
Applying the SGR~\cite{SGR} we can improve the accuracy of the
LT term in the large-$x$ region.
The resummed LT calculated at NLL deviates significantly from
the LL result, similarly to the NLO and LO at fixed order.
Unfortunately, next-to-next-to-leading log (NNLL) calculations
are not yet available to confirm that also resummed moments
converge at second order.
Nevertheless, the LT term is calculated to the best accuracy
currently available.

The extracted LT components of the proton and deuteron moments can be
combined to form moments of the neutron $F_2$ structure function.
In the nuclear Impulse Approximation (IA), the nuclear structure
function can be written as a convolution of the nucleon structure
function and a nucleon distribution function, $f^D$, in the deuteron.
In moment space this translates into a product of moments, so that
the neutron moments can be obtained from:
\begin{equation}\label{eq:nucl_cor}
M_n^n(Q^2)=\frac{2M_n^D(Q^2)}{N_n^D}-M_n^p(Q^2) ~,
\end{equation}
\noindent where $M_n^p$, $M_n^n$ and $M_n^D$ are the proton,
neutron and deuteron moments, respectively, and $N_n^D$ is the
moment of the function $f^D$.
The distribution function $f^D$ was calculated from various
deuteron wave functions \cite{osipenko_f2n}.

\begin{figure}[htb]
\vspace{9pt}
\includegraphics[bb=2cm 4cm 22cm 24cm, scale=0.4]{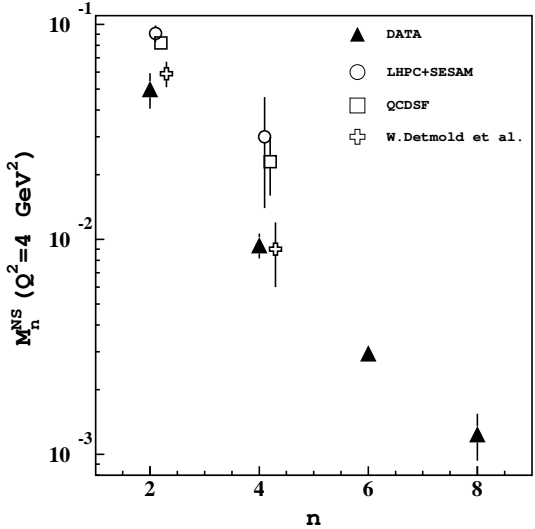}
\caption{Moments of the non-singlet $F_2$ structure function
compared with several lattice QCD simulations:
filled triangles - present analysis,
open circles - lattice simulations from Ref.\cite{Dolgov} with
		linear extrapolation,
open squares - lattice simulations from Ref.\cite{Goeckeler} with
		linear extrapolation,
open crosses - lattice simulations from Ref.\cite{Goeckeler}
		with chiral extrapolation from Ref.\cite{Detmold}.}
\label{fig:lattice}
\end{figure}

The extracted LT proton and neutron moments can be combined to form
Non-Singlet (NS) moments of the nucleon $F_2$ structure function,
which can then be compared to lattice QCD simulations.
A comparison of the extracted moments with recent lattice results
from several groups is shown in Fig.~\ref{fig:lattice}.
While a linear extrapolation of the lattice results to the physical
pion mass overestimates our data significantly for $n=2$ and 4,
the results extrapolated using chiral effective \cite{Detmold}
theory agree very well with our data.
The data for higher moments are also of high precision, and it would
be of considerable interest to compare these with higher lattice
moments, especially since the effects of chiral loops are expected
to be suppressed in the large-$n$ (large-$x$) domain. 

Another interesting result that can be obtained from the proton
and neutron moments is related to the behavior of $u$ and $d$
quarks in the proton in the $x \to 1$ limit.
At leading twist, the $d/u$ ratio at large $x$ can be extracted
directly from the ratio of the neutron to proton structure functions 
$F_2^n/F_2^p$, which is in turn related to the ratio of the moments
$M_n^n/M_n^p$ for large $n$.
Indeed, for $n \gg 1$ one finds that
$M_n^n/M_n^p (n \to \infty) = F_2^n/F_2^p (x \to 1)$.

\begin{figure}[htb]
\vspace{9pt}
\includegraphics[bb=2cm 4cm 22cm 24cm, scale=0.4]{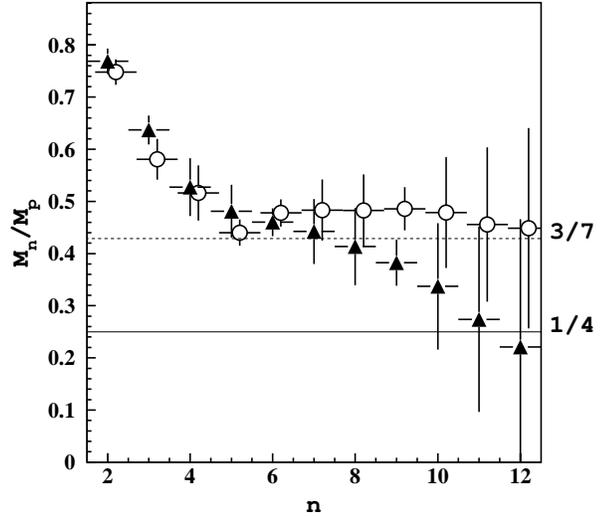}
\caption{Ratio of neutron to proton moments as a function of $n$:
filled triangles - this analysis,
open circles - ratio without nuclear corrections applied.
The solid (dashed) line indicates the scenario where $d/u \to 0 (1/5)$
for $x \to 1$ \protect\cite{MT}.}
\label{fig:nucl_cor}
\end{figure}

The $n/p$ structure function moment ratios are shown in
Fig.~\ref{fig:nucl_cor}.
Also indicated on the vertical axes are model predictions
for the $x \to 1$ limits, namely the ``standard'' 1/4 value
used in most of parton distribution fits (corresponding to 
a vanishing $d/u$ ratio), and the value 3/7 expected from
helicity conservation model (see Ref.~\cite{MT}).
The trend of our data is towards the lower
value of the model predictions with increasing $n$, although
the precision of the data does not exclude the higher value.

Indeed, at large $n$, both the data and the theoretical framework
become more problematic, making it more difficult to distinguish
between the different hypotheses.
From Fig.~\ref{fig:nucl_cor} one can also see the impact of the
nuclear corrections on the deuteron moments.
This is particularly evident for large $n$: for $n=12$, which
corresponds to $x$ values of around 0.75, the nuclear correction
introduces a factor of $\sim 2$.

\begin{figure}[htb]
\includegraphics[bb=2cm 4cm 22cm 24cm, scale=0.4]{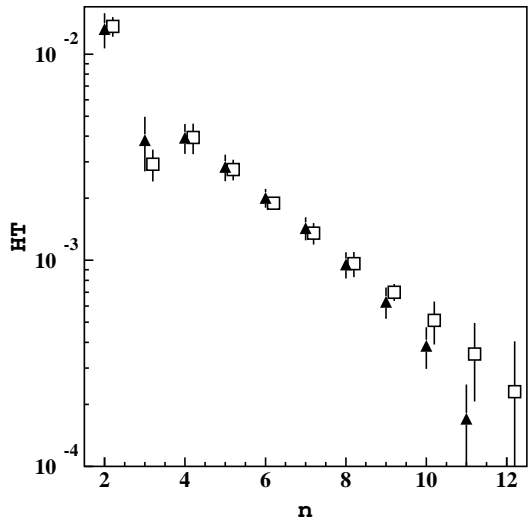}
\caption{Total higher twist contribution to the proton
(filled triangles) and deuteron (open squares) moments
at $Q^2=2$~(GeV/c)$^2$.}
\label{fig:hts}
\end{figure}

Once the LT contribution to moments is determined, one can
then study the isospin dependence of the HT contribution.
The HTs provide important information about multiparton
correlations inside the nucleon.
We assume that final state interactions and meson exchange
currents (in particular, the $1/Q^2$ components) is negligible
above $Q^2=1$~(GeV/c)$^2$, so that the same nuclear corrections
can be applied to the HTs as for the LT, Eq.~\ref{eq:nucl_cor}.
The total HT contributions to proton moments and the corrected
deuteron moments can then be compared, Fig.~\ref{fig:hts}.
This comparison indicates that the total HT contribution is
independent of isospin.
The isovector combinations $p-n$ of structure functions $F_2$
should therefore be free of HT contributions, within the
presented uncertainties.

In summary, we have analyzed experimental data on proton and deuteron
$F_2$ structure function in order to extract their moments, and 
performed an OPE analysis to separate leading and higher twist
contributions.
By combining proton and deuteron moments and applying nuclear 
corrections, we extracted moments of the neutron $F_2$ structure
function, paying particular attention to the issue of nuclear effects 
in the deuteron, which are increasingly important for higher moments.

The LT part of the non-singlet moments were then obtained and
related to lattice QCD moments available for $n=2$ and 4.
For large $n$, the ratio of the neutron to proton moments can
be related to the ratio of $u$ and $d$ quark contributions in
the proton in the $x \to 1$ limit.
The HT contribution is related to the physics beyond the
asymptotically free regime --- namely, multiparton correlations.
The results of our analysis can be summarized as follows:
\begin{itemize}
\item
the ratio of neutron to proton moments is consistent with
$F_2^n/F_2^p \to 1/4$ as $x \to 1$, although one cannot exclude
the higher value of 3/7 suggested by helicity conservation
arguments;
\item
the non-singlet moments are in excellent agreement with the
lattice data \cite{Dolgov,Goeckeler}, if these are extrapolated
to physical quark masses taking into account chiral loops
associated with the pion cloud \cite{Detmold}, but underestimate
the lattice results when linearly extrapolations are used;
\item
the total contribution of HTs is found to be isospin independent,
which implies that in the isovector combination $p - n$ of $F_2$
structure functions the HTs are consistent with zero.
\end{itemize}


\begin{thebibliography}{9}
\bibitem{osipenko_f2pd} M. Osipenko et al., Phys. Rev. D67 (2003) 092001;
Phys. Rev. C73 (2006) 045205.
\bibitem{SGR} S. Simula, Phys. Lett. B493 (2000) 325.
\bibitem{osipenko_f2n} M. Osipenko et al., Nucl. Phys. A766 (2006) 142.
\bibitem{Dolgov} D. Dolgov et al. (LHPC and SESAM Collaborations), Phys. Rev. D66 (2002) 034506.
\bibitem{Goeckeler} M. G\"{o}ckeler et al. (QCDSF Collaboration), Phys. Rev. D71 (2005) 114511.
\bibitem{Detmold} W. Detmold et al., Phys. Rev. Lett. 87 (2001) 172001.
\bibitem{MT} W. Melnitchouk and A.W. Thomas, Phys. Lett. B 377 (1996) 11.

\end{thebibliography}
\end{document}